\documentclass[preprintnumbers,article,amsmath,amssymb,floatfix,10pt,prd,twocolumn,
superscriptaddress,nofootinbib]{revtex4-2}
\usepackage{bm}
\usepackage{amsfonts}
\usepackage{latexsym}
\usepackage[latin1]{inputenc}
\usepackage{graphicx}
\usepackage{amsmath}
\usepackage{palatino}
\usepackage{mathpazo}
\usepackage{textcomp}
\usepackage{float}
\usepackage{booktabs}
\usepackage{dcolumn}
\usepackage{ragged2e}
\usepackage{hyperref}
\usepackage{tensor}
\usepackage[version=3]{mhchem}
\hypersetup{colorlinks,citecolor=blue}
\hypersetup{colorlinks=true,linkcolor=red,filecolor=magenta,    urlcolor=blue}
\usepackage{amsmath}
\usepackage{xcolor}
\usepackage{enumitem}
\usepackage{orcidlink}
\usepackage{epsfig}
\usepackage{subfigure}
\usepackage{commath}
\usepackage{caption}

\def\jnl@style{\it}
\def\aaref@jnl#1{{\jnl@style#1}}

\def\aaref@jnl#1{{\jnl@style#1}}

\def\aj{\aaref@jnl{AJ}}                   
\def\apj{\aaref@jnl{ApJ}}                 
\def\apjl{\aaref@jnl{ApJ}}                
\def\apjs{\aaref@jnl{ApJS}}               
\def\apss{\aaref@jnl{Ap\&SS}}             
\def\aap{\aaref@jnl{A\&A}}                
\def\aapr{\aaref@jnl{A\&A~Rev.}}          
\def\aaps{\aaref@jnl{A\&AS}}              
\def\mnras{\aaref@jnl{Mon.~Not.~Roy.~Astron.~Soc.}}             
\def\prd{\aaref@jnl{Phys.~Rev.~D}}        
\def\prc{\aaref@jnl{Phys.~Rev.~C}}  
\def\prl{\aaref@jnl{Phys.~Rev.~Lett.}}    
\def\qjras{\aaref@jnl{QJRAS}}             
\def\skytel{\aaref@jnl{S\&T}}             
\def\ssr{\aaref@jnl{Space~Sci.~Rev.}}     
\def\zap{\aaref@jnl{ZAp}}                 
\def\nat{\aaref@jnl{Nature}}              
\def\aplett{\aaref@jnl{Astrophys.~Lett.}} 
\def\apspr{\aaref@jnl{Astrophys.~Space~Phys.~Res.}} 
\def\physrep{\aaref@jnl{Phys.~Rep.}}      
\def\physscr{\aaref@jnl{Phys.~Scr}}       
\def\commat{\aaref@jnl{Comm.~Math.~Phys.}}              
\def\science{\aaref@jnl{Science}}               
\def\cqg{\aaref@jnl{Classical Quant.~Grav.}}            
\def\jpcs{\aaref@jnl{JPCS}}                                     
\def\ijmpd{\aaref@jnl{Int.~J.~Mod.~Phys.~D}}                    
\def\grg{\aaref@jnl{Gen.~Relat.~Gravit.}}               
\def\rpp{\aaref@jnl{Rep.~Prog.~Phys.}}          
\def\npa{\aaref@jnl{Nucl.~Phys.~A}}        
\def\lrr{\aaref@jnl{Living Rev.~Rel.}}                   
\def\jcap{\aaref@jnl{J.~Cosmology Astropart.~Phys.}}    
\def\rmp{\aaref@jnl{Rev.~Mod.~Phys.}}   
\def\epjc{\aaref@jnl{Eur.~Phys.~J.~C}} 
\def\plb{\aaref@jnl{~Phy.~Lett.~B}} 
\def\mpla{\aaref@jnl{Mod.~Phy.~Lett.~A}} 
\def\arxiv{\aaref@jnl{arxiv.org}}

\allowdisplaybreaks[1]

\date{\today}

\begin{document}

\title{Investigating early and late-time epochs in \texorpdfstring{$ f(Q) $}{Lg} gravity}

\author{Ameya Kolhatkar\orcidlink{0000-0002-9553-1220}}
\email{kolhatkarameya1996@gmail.com}
\affiliation{Department of Mathematics, Birla Institute of Technology and
Science-Pilani,\\ Hyderabad Campus, Hyderabad-500078, India.}

\author{Sai Swagat Mishra\orcidlink{0000-0003-0580-0798}}
\email{saiswagat009@gmail.com}
\affiliation{Department of Mathematics, Birla Institute of Technology and
Science-Pilani,\\ Hyderabad Campus, Hyderabad-500078, India.}

\author{P.K. Sahoo\orcidlink{0000-0003-2130-8832}}
\email{pksahoo@hyderabad.bits-pilani.ac.in}
\affiliation{Department of Mathematics, Birla Institute of Technology and
Science-Pilani,\\ Hyderabad Campus, Hyderabad-500078, India.}
\begin{abstract}
    \noindent In the following work, a new hybrid model of the form $ f(Q)=Q(1+a)+b\frac{Q_0^2}{Q} $ has been proposed and confronted using both early as well as late-time constraints. We first use conditions from the era of Big Bang Nucleosynthesis (BBN) in order to constrain the models which are further used to study the evolution of the Universe through the deceleration parameter. This methodology is employed for the hybrid model as well as a simple model of the form $ \alpha_1 Q+\alpha_2 Q_0 $ which is found to reduce to $\Lambda$CDM. The error bar plot for the Cosmic Chronometer (CC) and Pantheon+SH0ES datasets which includes the comparison with $\Lambda$CDM, has been studied for the constrained hybrid model. Additionally, we perform a Monte Carlo Markov Chain (MCMC) sampling of the model against three datasets -- CC, Pantheon+SH0ES, and Baryon Acoustic Oscillations (BAO) to find the best-fit ranges of the free parameters. It is found that the constraint range of the model parameter ($a$) from the BBN study has a region of overlap with the ranges obtained from the MCMC analysis. Finally, we perform a statistical comparison between our model and the $\Lambda$CDM model using AIC and BIC method.

\end{abstract}
\maketitle
\section{Introduction}
In a pursuit to make sense of the observed Universe, theory needs to catch up with the data. From the role of dark matter in structure formation to the cosmic expansion on the large scale, there is no theory that provides a complete answer. It is well understood that Einsteinian GR needs to be modified in some way to account for the large scale behaviour but also that it has to be made compatible with quantum mechanics to study systems with high energy densities like those found at the time of the very early Universe. While many approaches have been made over the years to resolve these issues, none of them has been cemented as the complete approach. Let us review some of these approaches. 

The Riemann tensor is at the heart of the General Theory of Relativity as first proposed by Albert Einstein in 1916. Thus the most natural way to extend or modify the theory without altering the nature of the matter content is by altering the geometric elements in it. The $ f(R) $ theory \cite{fR1,fR2} is a result of that. There are two more frameworks equivalent to GR -- The Teleparallel Equivalent to GR (TEGR) \cite{TEGR1,TEGR2,TEGR3} and the Symmetric Teleparallel Equivalent to GR (STEGR) \cite{STEGR,fQ} (More information about the equivalence between the different flavours of gravity can be found in \cite{grequiv} ). TEGR uses torsion ($ T $) based geometry instead of the curvature-based one and the STEGR uses geometry based on the non-compatibility of the metric tensor using the non-metricity scalar ($ Q $). Both of these approaches can be extended just like the GR case into $ f(T) $ and the $ f(Q) $ gravities respectively. There are also many other approaches to this like -- $ f(R,\mathcal{L_M})$ \cite{fRLm1,fRLm2,fRLm3,fRLm4}, $ f(R, T) $
\cite{fRT}, $ f(T, \mathcal{T})$\cite{harko,fTT2}, $f(Q, T)$ \cite{fQT} just to name a few. For our purposes, we shall focus on the extended STEGR or $f(Q)$ gravity. This theory has gained momentum in recent years with works on cosmology\cite{fQcosmo}, energy conditions \cite{fQEC}, cosmographical analysis \cite{fQcgraphy}, large scale structure \cite{fQLSS}, bouncing scenarios \cite{fQbounce}. An extensive review on $ f(Q) $ is given in \cite{fQreview}.

Now we will shift our focus to one of the initial events of the Universe which is known as Big Bang Nucleosynthesis (BBN). The time of occurrence of BBN is believed to be within a few minutes after the Big Bang. In 1939 Bethe \cite{bethe} introduced the idea of BBN which is also known as primordial nucleosynthesis. In this period, the Universe cooled down enough leading to the formation of light elements like \ce{^1 H}, \ce{^2 H}, \ce{^3 He}, \ce{^4 He}, \ce{^7 Li}. The relative abundance of \ce{^{4}He} in the universe is further explained by this phenomenon. The BBN constraints based on the current observational data on the primordial abundance of \ce{^{4}He} can be used to constrain any cosmological model. To explore many interesting findings from BBN constraints in the context of modified gravity, refer \cite{fQBBN,Capozziello,fTTbbn}. In this work, we have imposed the BBN constraints on a simple GR equivalent model and a new hybrid model.

Among the numerous models available for studying the evolution of the Universe, only those models that satisfy the observational constraints can be regarded as ``viable". The constrained models from BBN are tested against the observational datasets (CC, Pantheon+SH0ES, BAO). Also, the transition from deceleration to acceleration through the deceleration parameter $q(z)$ using those models is studied. We perform a statistical comparison between our model and the $\Lambda$CDM model using the AIC and BIC methods. This manuscript is organized in the following manner: Sec \ref{sec2} is devoted to the basic equations of $f(Q)$ gravity and two models are introduced in Sec \ref{sec3}. The BBN formalism has been discussed in Sec \ref{sec4} and the models are constrained in Sec \ref{sec5}. Sec \ref{sec6} \& \ref{sec7} are dedicated to test the models in different eras through deceleration parameter and observational datasets. Finally, we have concluded our work in Sec \ref{sec8}.

\section{\texorpdfstring{$ f(Q) $} {Lg} gravity and Cosmology}\label{sec2}
\noindent In the following section, we discuss the basics of $ f(Q) $ gravity and the resulting cosmology.

\noindent Any discussion on a modification of the conventional General Relativity starts with the modification in action. So in this case, we modify the Einstein-Hilbert ($ S_{EH} $) action as follows 
\begin{equation}\label{fQaction}
    S_{EH}\longrightarrow S_Q=\int d^4x\sqrt{-g}\Big( -\frac{f(Q)}{16\pi G}+\mathcal{L}_M \Big)
\end{equation}
where $ \mathcal{L}_M $ is the matter Lagrangian, $ Q=Q_{\alpha\beta\gamma}P^{\alpha\beta\gamma} $ is the non-metricity scalar that arises from the non-metricity tensor $ Q_{\alpha\beta\gamma}=\nabla_\alpha g_{\beta\gamma} $ and the superpotential tensor $ P^{\alpha\beta\gamma} $ which is defined as
\begin{equation}\label{superpotential}
    P^\alpha_{\mu\nu}=-\frac{1}{2}L^\alpha_{\mu\nu}+\frac{1}{4}(Q^\alpha-\Tilde{Q}^\alpha)g_{\mu\nu}-\frac{1}{4}\delta^\alpha_{(\mu}Q_{\nu)}
\end{equation}
with the disformation tensor $ L^\alpha_{\mu\nu}=\frac{1}{2} 
\tensor{Q}{^\alpha_{\mu\nu}}-\tensor{Q}{_{(\mu\nu)}^\alpha} $ and the two independent traces being $ Q_\alpha=g^{\mu\nu}Q_{\alpha\mu\nu} $ and $ \Tilde{Q}_\alpha=g^{\mu\nu}Q_{\mu\alpha\nu} $. A variation of the action \eqref{fQaction} with respect to the metric yields the following field equations
\begin{multline}\label{fQfieldequations}
    \frac{2}{\sqrt{-g}}\nabla_\alpha(\sqrt{-g}f_Q\tensor{P}{^\alpha_{\mu\nu}})-\frac{1}{2}g_{\mu\nu}f+f_Q(P_{\mu\alpha\beta}\tensor{Q}{_\nu^{\alpha\beta}}\\-2Q_{\alpha\beta\mu}\tensor{P}{^{\alpha\beta}_\nu})=8\pi G T_{\mu\nu}
\end{multline}
where $ f_Q\equiv\frac{df}{dQ} $ and $ T_{\mu\nu} $ is the energy-momentum tensor defined as 
\begin{equation}\label{energymomentumtensor}
    T_{\mu\nu}=-\frac{2}{\sqrt{-g}}\frac{\delta(\sqrt{-g}\mathcal{L}_M)}{\delta g^{\mu\nu}}
\end{equation}
In order to proceed, we must choose a specific metric. Here we choose the spatially flat FLRW (Friedmann-Robertson-Le-Maitre-Walker) metric given by the infinitesimal line element
\begin{equation}\label{FLRWlineelement}
    ds^2=-dt^2+a^2(t)(dx^2+dy^2+dz^2)
\end{equation}
Here, $a(t)$ is the scale factor. The field equations \eqref{fQfieldequations} in this case take the form
\begin{gather}\label{FriedmannequationsfQ}
    6f{_Q}H^2-\frac{1}{2}f=8\pi G\rho\\
    (12H^2f_{QQ}+f_Q)\Dot{H}=-4\pi G(\rho+p)
\end{gather}
where $ f_{QQ}=\frac{d^2f}{dQ^2} $, $ H(t)=\frac{\Dot{a}}{a} $ is the Hubble parameter and the overdot represents derivative with respect to time. Note that $ \rho=\rho_m+\rho_r $ and $ p=p_m+p_r $. One can modify the equations in order to get 
\begin{gather}\label{friedmann}
    3H^2=8\pi G(\rho+\rho_Q),\\
     \Dot{H}=-4\pi G(\rho+p+\rho_Q+p_Q)
\end{gather}
where $ \rho_Q $ and $ p_Q $ are identified with
\begin{gather}\label{rhoQpQ}
    \rho_Q= \frac{1}{16\pi G}\left( Q(1-2f_Q)+f \right),\\  p_Q= \frac{1}{16\pi G}\left( 4\Dot{H}(f_Q-1)-f+Q(8f_{QQ}\Dot{H}+2f_Q-1) \right).
\end{gather}
Let us rewrite the Friedmann equation in a form that will be useful in the later sections.
\begin{equation}\label{friedmanninQ}
    \frac{Q}{Q_0}=\Omega(z)+\Omega_Q.
\end{equation}
Here, $ \Omega(z)=\Omega_{m0}(1+z)^3+\Omega_{r0}(1+z)^4 $, $ \Omega_Q=\frac{8\pi G}{3H_0^2}\rho_Q $ and $ Q_0=6H_0^2 $.

We now turn to the technique of cosmography to calculate the deceleration parameter $ q(z) $. This technique is a completely model independent approach that works by Taylor expanding the scale factor around the present time as follows
\begin{equation}\label{Taylorexpansionofa}
    a(t)=a(t_0)\left(1+H_0\Delta t-\frac{1}{2}q_0H_0^2\Delta t^2+...\right)
\end{equation}
where the cosmographical parameters at an arbitrary time $ t $ can be written as
\begin{enumerate}
    \item $ H(t) $ is the Hubble parameter with $ 
H(t)=\frac{\Dot{a}}{a} $.
    \item $ q(t) $ is the deceleration parameter with $ q=-\frac{1}{a}\frac{d^2a}{dt^2}H^{-2} $.
\end{enumerate}
It is easy to see that $ q(t) $ can be written in terms of the Hubble parameter so that
\begin{equation}\label{qina}
    q(t)=-1-\frac{\Dot{H}}{H^2}.\quad
\end{equation}
Further, one can use the scale factor-redshift relation and rewrite \eqref{qina} in terms of $ z $. 
\begin{equation}\label{ainz}
    a(t)=\frac{1}{1+z}\Longrightarrow\frac{d}{dt}\longrightarrow-(1+z)H(z)\frac{d}{dz}
\end{equation}
\begin{equation}\label{qinz}
    q(z)=-1+(1+z)\frac{H'(z)}{H(z)}.
\end{equation}
Where the prime ($ ' $) denotes the derivative with respect to $ z $.
\section{The models}\label{sec3}
In the following section, we discuss the models used and their physical motivations.
\subsection{The GR equivalent : \texorpdfstring{$ f(Q)=\alpha_1 Q+\alpha_2 Q_0 $}{Lg}}
This is the simplest model that reduces to GR in the limit $ \alpha_1=1 $ and $ \alpha_2=1-\Omega_0\approx 0.7 $, $\Omega_0$ is the present value of $\Omega$. The corresponding expression for $ \rho_Q $ is
\begin{equation}\label{rhoQmodel1}
    \rho_Q=\frac{Q_0}{16\pi G}\left( \alpha_2+(1-\alpha_1)\frac{Q}{Q_0} \right).
\end{equation}
 This can be inserted into \eqref{friedmanninQ} to get
\begin{equation}\label{Hzmodel1}
    H(z)=H_0\sqrt{\frac{1}{\alpha_1}\left( \Omega(z)+\alpha_2 \right)}.
\end{equation}
At $ z=0 $, we find $ \alpha_2=\alpha_1-\Omega_0 $. Using \eqref{Hzmodel1} in \eqref{qinz}, we get
\begin{equation}\label{qinzmodel1}
    q(z)=-1+\frac{(1+z)\Omega'(z)}{2(\Omega(z)+\alpha_2)}.
\end{equation}
\subsection{The Hybrid Model : \texorpdfstring{$ f(Q)=Q(1+a)+b\frac{Q_0^2}{Q} $}{Lg}}
We propose this model in order to take into account the effects from both the early and late-time epochs of the Universe. With $ Q\equiv Q(z)=6H^2(z) $, one can make the following observation. At very high redshift values (the early Universe), $ H(z) $ varies as the square of temperature and hence takes higher values which makes terms with inverse powers of $ Q $ vanish. On the other hand, at lower redshifts (the future), $ H(z) $ takes lower values which makes the positive powers of $ Q $ insignificant. In order to get the best of both these epochs, we propose the hybrid model which is the simplest possible combination of the linear and inverse power terms in $ Q $.

This model reduces to GR when the dimensionless parameters $ a,b $ both take the value $ 0 $. The corresponding expression for $ \rho_Q $ is 
\begin{equation}\label{rhoQmodel2}
    \rho_Q=\frac{1}{16\pi G}\left( -a Q+3b\frac{Q_0^2}{Q} \right).
\end{equation}
Thus, \eqref{friedmanninQ} takes the form
\begin{equation}\label{Hzmodel2}
    H(z)=H_0\sqrt{\frac{\Omega(z)+\sqrt{\Omega^2(z)+12b(1+a)}}{2(1+a)}}.
\end{equation}
Substituting $ z=0 $, we find $ 3b=a+\Omega_{Q0} $ with $ \Omega_{Q0}=1-\Omega_0\approx 0.7 $. Correspondingly, we find that the expression for the deceleration parameter, using \eqref{qinz} becomes
\begin{equation}\label{qinzmodel2}
    q(z)=-1+\frac{(1+z)\Omega'(z)}{\sqrt{\Omega^2(z)+4(a+\Omega_{Q0})(a+1)}}.
\end{equation}
\section{Constraints from Big Bang Nucleosynthesis}\label{sec4}
In the following section, we discuss the BBN era in the $ f(Q) $ gravity framework. The first point to note is that BBN takes place during the radiation-dominated era, and so we have $ a(t)\sim t^{1/2} $ and $ H(t)\sim\frac{1}{2t} $. Further, the first Friedmann equation takes the form $ 3H^2=8\pi G\rho_R $ where $ \rho_R $ accounts for the energy density of the relativistic particles given by
\begin{equation}\label{rhoR}
    \rho_R=\frac{\pi^2 g_* T^4}{30}.
\end{equation}
Here, $ g_*\approx 10 $ and $ T $ are the effective number of degrees of freedom and the corresponding temperature respectively. Introducing the definition of the reduced Planck mass where $ M_p=\frac{1}{\sqrt{8\pi G}}=1.22\times 10^{19} GeV $\footnote{$ M_{pl}=\sqrt{8\pi} M_p $ is the Planck mass.} we can replace the factors of $ 8\pi G $. It is to be noted that we shall use the deviations in the expression for the freeze-out temperature resulting from the modification of GR and hence we label the Hubble parameter corresponding to the later with $ 
H_{GR} $. Hence,
\begin{equation}\label{HGR}
    H=\sqrt{\frac{\rho_R}{3M_p^2}}\equiv H_{GR}.
\end{equation}
Inserting \eqref{rhoR} into \eqref{HGR} the expression of the Hubble parameter in terms of temperature can be obtained as
\begin{equation}\label{HinT}
    H(T)=\sqrt{\frac{\pi^2 g_*}{90 M_p^2}}T^2.
\end{equation}
We see that \eqref{friedmann} can be written in terms of $ H_{GR} $ as follows
\begin{equation}
    H=H_{GR}\sqrt{1+\frac{\rho_Q}{\rho_R}}
\end{equation}
and since $ \frac{\rho_Q}{\rho_R}<<1 $ in the radiation era, 
\begin{equation}\label{DeltaH}
    \Delta H\approx\frac{\rho_Q}{2\rho_R}H_{GR}.
\end{equation}
Two parameters that are important in studying the Universe are the neutron-to-proton ratio and the baryon-to-photon ratio of which we shall focus on the prior. The neutron-to-proton ratio, before BBN ($ T>>1 MeV $), was $ 1:1 $ due to the weak interaction reactions being in equilibrium. Neutrons and protons convert into each other through three reactions (i) $ n\longrightarrow e^- +p^+ +\Bar{\nu}_e $, (ii) $ n+\nu_e\longrightarrow e^- +p^+ $ and (iii) $ n+ e^+\longrightarrow p^+ + \Bar{\nu}_e $. Around this time, the expansion rate was much less as compared to the rate of the reactions. When temperatures dropped, however, to around $ T=0.7 MeV $, the reaction rate slowed down and the expansion rate overtook the same. This caused the neutron-to-proton ratio to ``freeze out" at around $ 1:6 $. But since free neutrons are unstable, the unfused ones decayed further into protons which finally set the neutron-to-proton ratio at $ 1:7 $. The neutrons that did fuse, became the \ce{^4 He} nuclei which is why, the mass fraction of this nucleus is the standard quantity to study BBN. The primordial mass fraction of \ce{^4 He} is expressed as
\begin{equation}\label{Yp}
    Y_p=e^{-(t_n-t_f)/\tau}\frac{2x(t_f)}{1+x(t_f)}
\end{equation}
where $ t_f $ is the freeze-out time , $ t_n $ is the freeze-out time corresponding to BBN, $ \tau $ is the mean lifetime of a neutron and $ x(T_f)=e^{-\mathcal{Q}/T(t_f)} $ with $ \mathcal{Q}=m_n-m_p=1.29\times 10^{-3} GeV $.
Denoting with $ \lambda_{np}(T) $ and $ \lambda_{pn}(T) $ the conversion rates for neutrons decaying into protons and vice-versa respectively, we can find the total conversion rate $ \lambda_{tot}(T)=\lambda_{np}(T)+\lambda_{pn}(T) $ as follows
\begin{equation}
    \lambda_{tot}(T)=4AT^3(4!T^2+2\times 3!\mathcal{Q}T+2!\mathcal{Q}^2)
\end{equation}
where $ A=1.02\times 10^{-11} GeV^{-4} $. Since $ \mathcal{Q}<<1 $, $ \lambda_{tot}(T)\approx c_q T^5 $ with $ c_q=4A4!=9.8\times 10^{-10} GeV^{-4} $. Using this along with the fact that the expansion rate was almost the same as the rate of weak interactions around the freeze-out temperature, $ H(T_f)=\lambda_{tot}(T_f) $ gives us
\begin{equation}\label{freezeoutT}
    T_f=\left(\frac{\pi^2 g_*}{90 M_p^2 c_q^2}\right)^{\frac{1}{6}}\approx 0.0006 GeV.
\end{equation}
Furthermore, since $ H_{GR}\approx c_q T^5 $, $ \Delta H_{GR}\approx 5c_q T^4\Delta T $. Thus, with \eqref{DeltaH} and $ T=T_f $
\begin{equation}\label{BBNTf}
    \frac{\Delta T_f}{T_f}=\frac{\rho_Q}{\rho_R}\frac{H_{GR}}{10c_q T_f^5}.
\end{equation}
A similar relation of the fractional deviation in the mass fraction can be obtained as
\begin{equation}\label{deltaYp}
    \frac{\Delta Y_p}{Y_p}=\left[ \left(1-\frac{Y_p}{2\lambda}\right)ln\left(\frac{2\lambda}{Y_p}-1\right)-\frac{2t_f}{\tau} \right]\frac{\Delta T_f}{T_f}.
\end{equation}
Here, $ \lambda=e^{-(t_n-t_f)/\tau} $ and from \cite{fQBBN}, $ Y_p=0.2476 $ and $ |\Delta Y_p|<10^{-4} $. Thus from \eqref{deltaYp}, the estimate for the bound for $ \frac{\Delta T_f}{T_f} $ to be
\begin{equation}\label{BBNTfobs}
     \bigg|\frac{\Delta T_f}{T_f}\bigg| < 4.7\times 10^{-4}.
\end{equation}
For a detailed discussion on deriving the BBN parameters, refer to \cite{fQBBN}.
\section{BBN constraints for the $ f(Q) $ models}\label{sec5}
\subsection{The GR equivalent : \texorpdfstring{$ f(Q)=\alpha_1 Q+\alpha_2 Q_0 $}{Lg}}
\begin{figure}[H]
    \centering
    \includegraphics[scale=0.7]{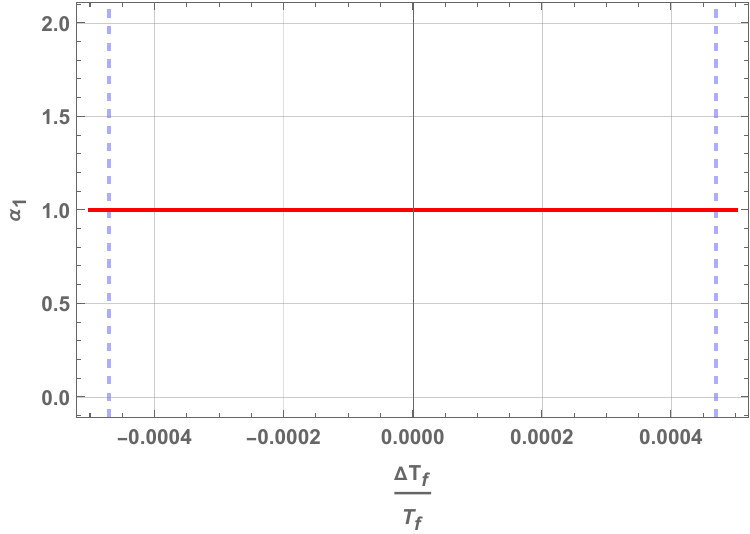}
    \caption{The fractional deviation in the freeze-out temperature vs the free model parameter.}
    \label{BBN-alpha1-model1}
\end{figure}

Although it is well known that this model reduces to STEGR in the limit that $ \alpha_1=1 $ and $ \alpha_2\equiv\Lambda\approx0.7 $, it is good to conduct this analysis and derive the conditions on the dimensionless free parameters $ 
\alpha_1,\alpha_2 $ to match them with our expectations. Using \eqref{friedmanninQ}, the Hubble parameter can be expressed as
\begin{equation}
    H(z)=H_0\sqrt{\frac{1}{\alpha_1}\left( \Omega(z)+\alpha_2 \right)}.
\end{equation}
Putting $ z=0 $ yields $ \alpha_2=\alpha_1-\Omega_0 $ with $ \Omega_0=\Omega_{m0}+\Omega_{r0}\approx0.30005 $. The energy density corresponding to $ Q $ in the radiation-dominated era is 
\begin{equation}\label{rhoQmodel1R}
    \rho_Q=\frac{1}{2}Q_0M_p^2\left( \alpha_1-\Omega_0+\frac{2\rho_R(1-\alpha_1)}{Q_0M_p^2} \right).
\end{equation}
Hence, using \eqref{rhoQmodel1R} along with \eqref{BBNTf} and \eqref{BBNTfobs}, the expression for the fractional deviation in the freeze-out temperature can be plotted against the free model parameter $ \alpha_1 $ (see \autoref{BBN-alpha1-model1}).
Notice that the free model parameter takes a constant value of $ \alpha_1=1 $ throughout the range of the fractional deviation. This immediately sets the other free parameter to take a value of $ 
\alpha_2\approx 0.7 $. This reduces the model to take the following form for the Hubble parameter
\begin{equation}
    H(z)=H_0\sqrt{\Omega_{m0}(1+z)^3+\Omega_{r0}(1+z)^4+0.7}
\end{equation}
which is nothing but the $ \Lambda $CDM model where $ \Lambda\approx0.7 $.

\subsection{Hybrid Model : \texorpdfstring{$ f(Q)=Q(1+a)+\frac{b Q_0^2}{Q} $}{Lg}}
\begin{figure}[H]
    \centering
    \includegraphics[scale=0.7]{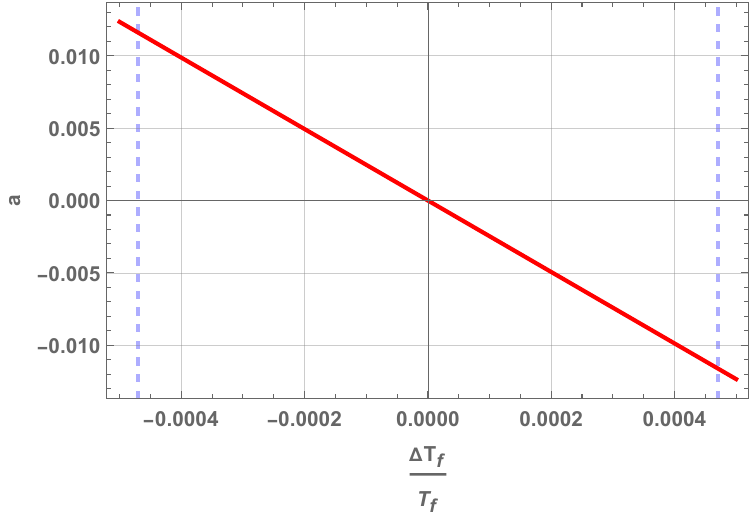}
    \caption{The parameter $ a $ plotted against the deviation in the freeze-out temperature.}
    \label{BBN-beta1-model2}
\end{figure}

The Friedmann equation \eqref{friedmanninQ} for this model takes the form 
\begin{equation}
    u=\Omega(z)+\left( -a u+\frac{3b}{u} \right)   
\end{equation}
where $u=Q/Q_0$. Realizing that the variable $ u \propto H^2(z) $, we discard one of the roots with the negative sign. Also using $ 3b=a+\Omega_{Q0} $, we get 
\begin{equation}
    H(z)=H_0\sqrt{\frac{\Omega(z)+\sqrt{\Omega^2(z)+4(a+\Omega_{Q0}) (1+a)}}{2(1+a)}}. 
\end{equation}
We can now calculate the expression for $ \rho_Q $ in the radiation-dominated era which comes out to be
\begin{equation}\label{rhoQmodel2R}
    \rho_Q=-a\rho_r+\left(a+\Omega_{Q0}\right)\frac{Q_0^2 M_p^4}{4\rho_r}.
\end{equation}
Constraining the free parameter $a$ using \eqref{rhoQmodel2R} with \eqref{BBNTf} and \eqref{BBNTfobs}, we obtain the range $ a\in[-0.0115907,0.0115907]$ in \autoref{BBN-beta1-model2}.
\section{Evolution from the deceleration parameter}\label{sec6}
In this section, we plot the expressions for the deceleration parameters against redshift for both the models which have been constrained with BBN to observe the evolution of the Universe. A transition form decelerating to accelerating phase is expected as predicted by the $ \Lambda $CDM model. 
\subsection{The GR equivalent}
\begin{figure}[H]
    \centering
    \includegraphics[scale=0.7]{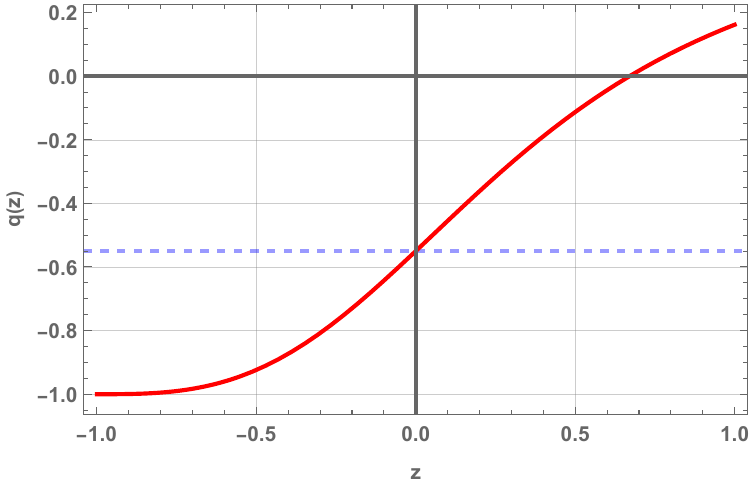}
    \caption{The deceleration parameter vs redshift for $ \alpha_1=1 $. The blue dashed line corresponds $ q_0=-0.5499 $.}
    \label{qz-model1}
\end{figure}

Plotting \eqref{qinzmodel1} against $ z $ in \autoref{qz-model1}, we depict that the Universe goes from a decelerating phase to an accelerating phase around $ z=0.7 $, which is the most accepted value for transition redshift.

\subsection{The Hybrid Model}
\begin{figure}[H]
    \centering
    \includegraphics[scale=0.54]{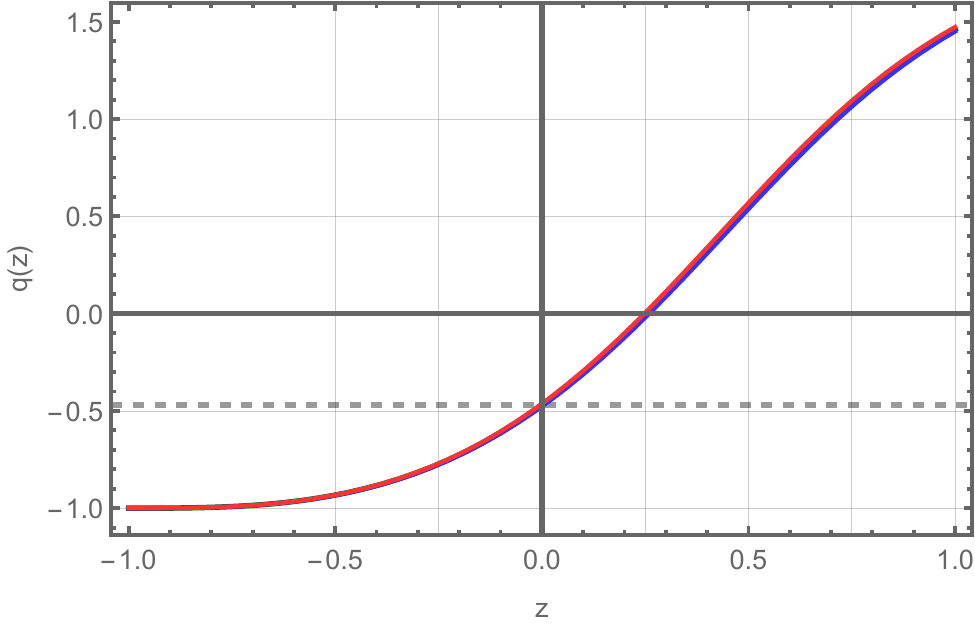}
    \includegraphics[scale=0.54]{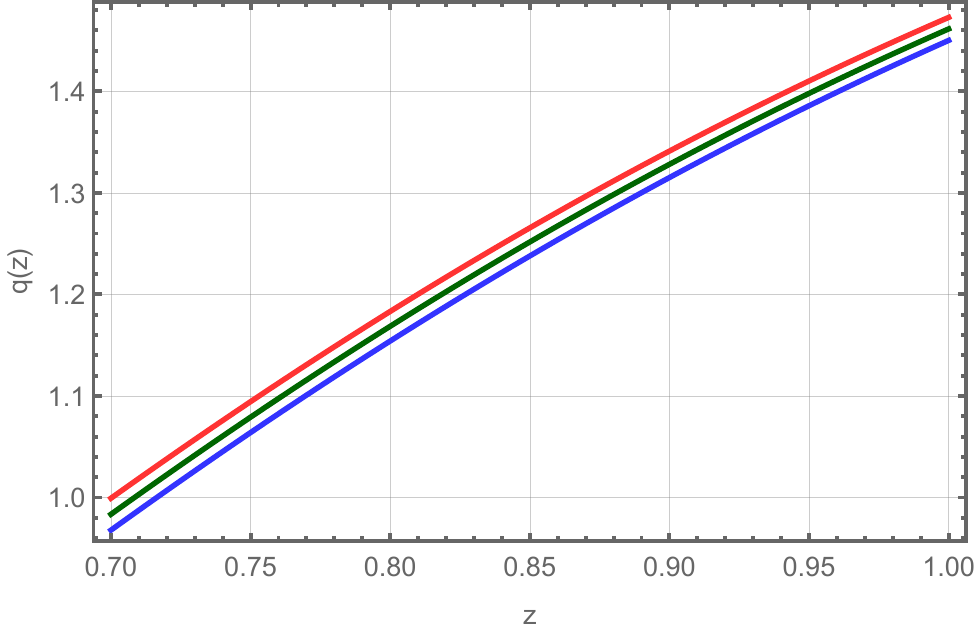}
    \caption{The deceleration parameter vs redshift for three values of $ a $ with red curve denoting $ a=-0.0115907 $, the green curve with $ a=0
 $ and the blue curve denoting $ a=0.0115907 $. The orange dashed line corresponds to $ q_0=-0.471 $.}
    \label{qz-model2}
\end{figure}

Plotting \eqref{qinzmodel2} against $ z $ in \autoref{qz-model2}, we find the transition redshift is around $ z=0.255 $. Although this value is quite low as compared to the widely accepted value of around $ 
z=0.7 $, this falls in the range of the study conducted in \cite{transitionredshift}. Notice that all three different curves merge shortly after the transition to the acceleration phase. 

\section{Data Analysis}\label{sec7}
\subsection{Cosmic Chronometer (CC) dataset}
The Hubble parameter $ H(z) $ is considered to be one of the most important cosmological parameters to measure the rate of expansion of the Universe. This can be expressed in terms of redshift ($z$) and time ($t$) as
\begin{equation}
    H(z)=-\frac{1}{1+z} \frac{dz}{dt}
\end{equation}.
Since $dz$ is obtained from a spectroscopic survey, $dt$ can be used to determine the model-independent value of the Hubble parameter. The Cosmic Chronometer method is adopted because of its ability to measure the H value without any cosmological assumptions. In the CC method, 31 data points have been used which are obtained from various sources \cite{Jimenez,
Simon, Stern, Moresco} with a redshift range varying from 0.07 to 2.42 (see the Hubble values for each redshift \cite{fTT2}). In this work, the MCMC analysis has been performed using the Chi-square function
\begin{equation}
    \chi^2_{CC}=\sum_{i=1}^{31}\frac{[H_i^{th}(\theta_s,z_i)-H_i^{obs}(z_i)]^2}{\sigma^2_{CC}(z_i)}.
\end{equation}

\begin{itemize}
    \item[] $ H_i^{th} $ -- Theoretical Hubble parameter value
     \item[] $ H_i^{obs} $ -- Observed Hubble parameter value
    \item[] $ \theta_s $ -- Cosmological background parameter space
    \item[] $ \sigma_{CC} $ -- Standard error in observed values
\end{itemize}
To obtain the best-fit range of our parameters in Fig. \ref{fig6}, we have used 100 walkers and 1000 steps in our analysis. Also, the prior range for the Hubble parameter is taken as $(60,85)$, for density parameter $\Omega_{m0}$ as $(0,1)$, and the range obtained from the BBN constraints is taken as prior for the model parameter $a$.

\subsection{Type Ia supernovae(SNe Ia)}
The original Pantheon sample has been upgraded by increasing the sample size with the addition of multiple cross-calibrated photometric systems of SNe and a wider range of redshift. The Pantheon+ analysis \cite{brout1,
brout2,scolnic} has been deemed revolutionary, in the context of cosmic evolution. In this work, we have used the Pantheon+SH0ES sample which consists of  1701 light curves of 1550 distinct Type Ia supernovae with a redshift range from 0.00122 to 2.26137. For the MCMC analysis, the chi-square function is defined as,
\begin{equation}
    \chi^2_{SN}=\sum_{i,j=1}^{1701} \nabla \mu_i \,\ (C_{SN}^{-1})_{ij} \,\ \nabla \mu_j
\end{equation}
where $\nabla \mu_i=\mu_i^{th}(z_i,\theta)-\mu_i^{obs}$ is the difference between theoretical and observational distance modulus.

\begin{itemize}
    \item[] $ \mu_i^{th} $ -- Theoretical distance modulus
     \item[] $ \mu_i^{obs} $ -- Observed distance modulus
    \item[] $ \theta $ -- Parameter space
    \item[] $ C_{SN} $ -- Covariance matrix
\end{itemize}

Further one can calculate the theoretical distance modulus using the formula,
\begin{equation}
    \mu_i^{th}(z,\theta)=  \mathrm{5 \,\ log} \,\ D_l(z, \theta)+25
\end{equation}
where $D_l(z,\theta)$ is the luminosity distance, defined as
\begin{equation}
    D_l(z,\theta)= (1+z) \int_0^z \frac{dx}{H(x)}.
\end{equation}
The above formulas along with the observational values and the same priors as CC methods are used to run the MCMC analysis which can be found in Fig. \ref{fig6}.
\subsection{Baryonic Acoustic Oscillations (BAOs)}

A collection of surveys from the 6-degree Field Galaxy Survey, Sloan Digital Sky Survey, and WiggleZ Dark Energy Survey\cite{blake} compose the Baryonic Acoustic Oscillation (BAO) data set measured at 6 distant redshifts. The sound horizon $(r_s)$ is regulated by the BAO observations and can be used to measure distances and the Hubble parameter at the corresponding redshifts. It is visible at the photon decoupling epoch with redshift $z_*$ and defined as 
\begin{equation}
    r_s(z_*)=\frac{c}{\sqrt{3}} \int_0^{\frac{1}{1+z_*}} \frac{(a^2 H)^{-1} da }{\sqrt{ 1+(3 \Omega_{b0}/4 \Omega_{\gamma 0})a}}
\end{equation}
where $c$, $\Omega_{b0}$, and $\Omega_{\gamma 0}$ denote the speed of light, present baryon, and photon densities, respectively. To obtain the BAO constraints $\frac{d_A(z_*)}{D_v(z_{BAO})}$ is used \cite{Percival,Beutler} and $z_*$ is considered to be 1091. Here $d_A(z_*)$ and $D_v(z_{BAO})$ are the angular distance and dilation scale, respectively. They are defined as follows,
\begin{gather}
    d_A(z)=\int_0^z \frac{dz'}{H(z')}\\
    D_v(z)= \left(\frac{d_A(z)^2 c z}{H(z)}\right)^{1/3}
\end{gather}

For MCMC analysis, the same priors, steps, and walkers as in the CC dataset are used. The chi-square function for BAO is defined as
\begin{equation}
   \chi^2_{BAO}= X^T C^{-1} X
\end{equation}
where $X$ and $C^{-1}$ \cite{Giostri} are ,

\begin{widetext}

$X=\begin{bmatrix}
\frac{d_A(z_*)}{D_v(0.106)} - 30.95\\
\frac{d_A(z_*)}{D_v(0.2)} - 17.55\\
\frac{d_A(z_*)}{D_v(0.35)} - 10.11\\
\frac{d_A(z_*)}{D_v(0.44)} - 8.44\\
\frac{d_A(z_*)}{D_v(0.6)} - 6.69\\
\frac{d_A(z_*)}{D_v(0.73)} - 5.45
\end{bmatrix}\medskip$ \\
$C^{-1}=\begin{bmatrix}
0.48435 & -0.101383 & -0.164945 & -0.0305703 & -0.097874 & -0.106738\\
-0.101383 & 3.2882 & -2.454987 & -0.0787898 & -0.252254 & -0.2751\\
-0.164945 & -2.454987 & 9.55916 & -0.128187 & -0.410404 & -0.447574\\
-0.0305703 & -0.0787898 & -0.128187 & 2.78728 & -2.75632 & 1.16437\\
-0.097874 & -0.252254 & -0.410404 & -2.75632 & 14.9245 & -7.32441\\
-0.106738 & -0.2751 & -0.447574 & 1.16437 & -7.32441 & 14.5022\\
\end{bmatrix}$

\end{widetext}

\begin{widetext}

\begin{figure}[H]
\caption{Comparision of the Hybrid model $f(Q)=Q(1+\beta_1)+\beta_2\frac{Q_0^2}{Q} $ with the Hubble and Pantheon+SHOES dataset along with the $\Lambda$CDM model.}\label{errorbar}
  \centering
  \subfigure[Error bar plot of 31 points of Hubble dataset. The red curve and blue dotted curve represent the Hubble function for the Hybrid model and the $\Lambda$CDM model, respectively.]{\includegraphics[scale=0.6]{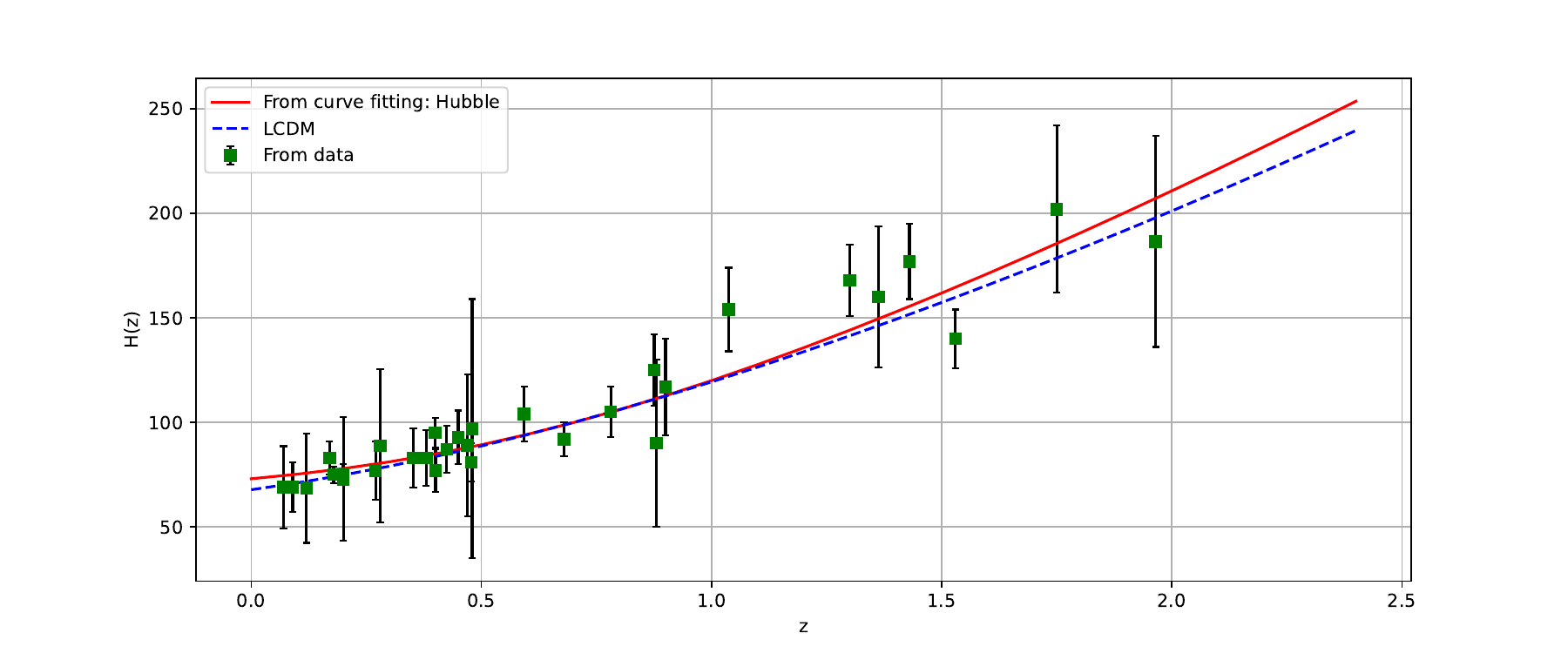}}\hfill \vspace{0.8 in}
  \subfigure[Error bar plot of 1701 points of Pantheon+SHOES dataset. The red curve and blue dotted curve represent the distance modulus function for the Hybrid model and the $\Lambda$CDM model, respectively.]{\includegraphics[scale=0.6]{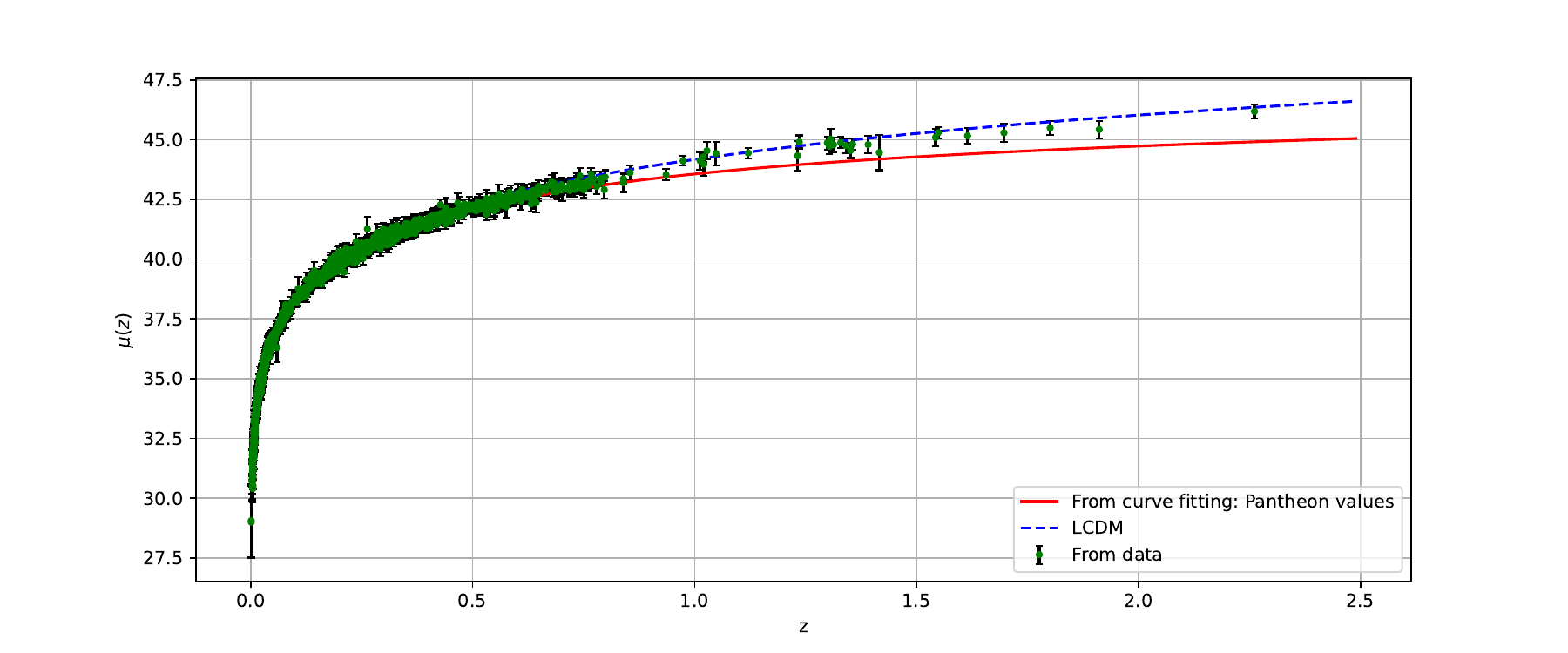}}
\end{figure}
\end{widetext}

\begin{widetext}

\begin{figure}[H]
\caption{MCMC analysis to find the best-fit range for the free parameters using CC, PANTHEON+SH0ES, and BAO samples.}
 \label{fig6}
  \centering
  \subfigure[Constraints on the parameters $H_0$, $\Omega_{m0}$ and $a$ using CC sample. The dark shaded region represents the $1\sigma$ $(68\%)$ confidence level and the light shaded region represents the $2\sigma$ $(95\%)$ confidence level.]{\includegraphics[scale=0.69]{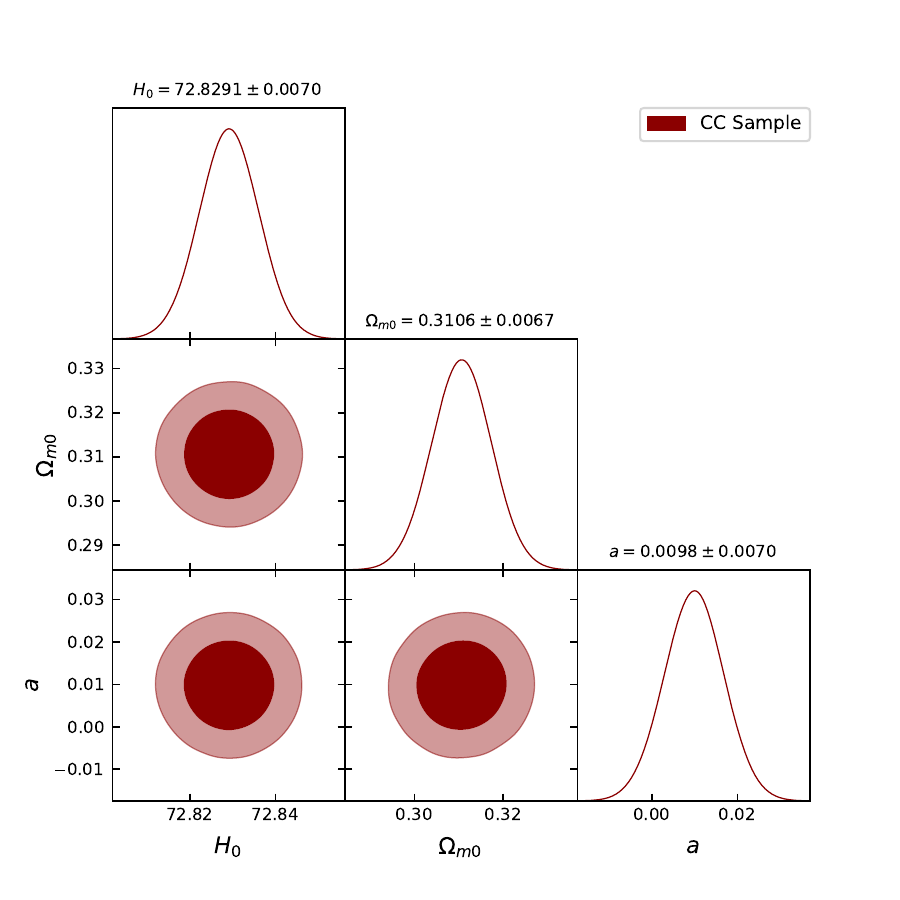}}\hfill
  \subfigure[Constraints on the parameters $H_0$, $\Omega_{m0}$ and $a$ using PANTHEON+SH0ES sample. The dark shaded region represents the $1\sigma$ $(68\%)$ CL and the light shaded region represents the $2\sigma$ $(95\%)$ CL. ]
  {\includegraphics[scale=0.69]{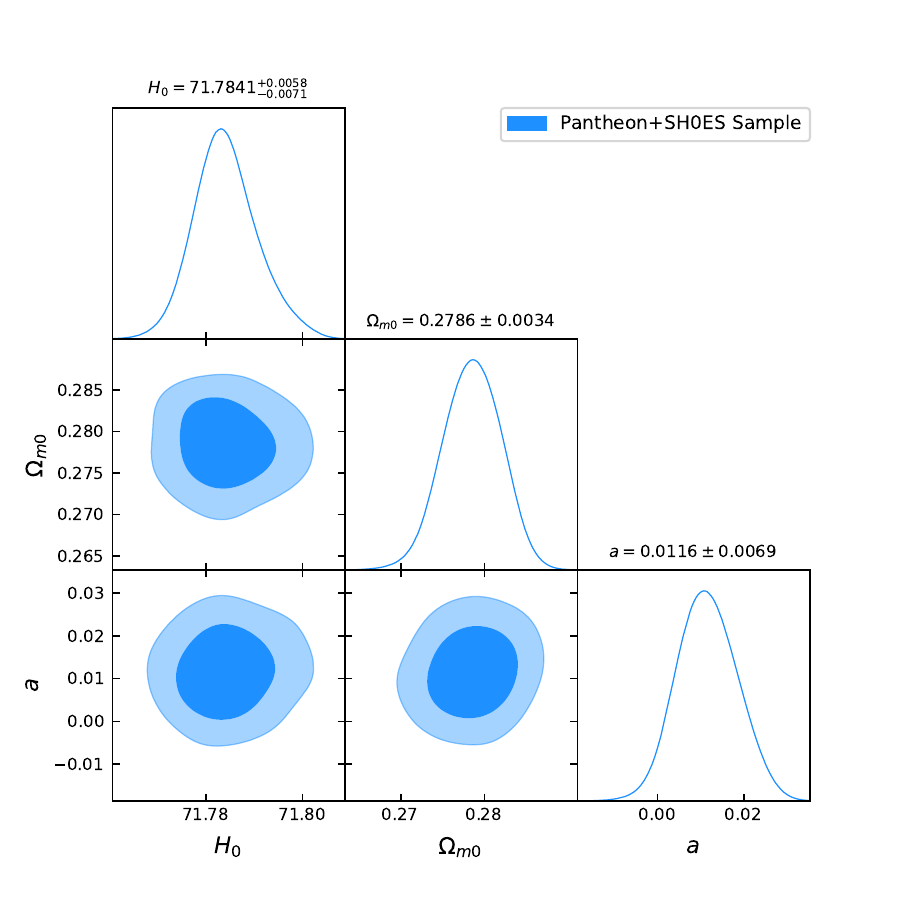}}\hfill
  \end{figure}%
\begin{figure}[ht]\ContinuedFloat
    \centering
  \subfigure[Constraints on the parameters $H_0$, $\Omega_{m0}$ and $a$ using BAO sample. The dark shaded region represents the $1\sigma$ $(68\%)$ CL and the light shaded region represents the $2\sigma$ $(95\%)$ CL. ]
  {\includegraphics[scale=0.69]{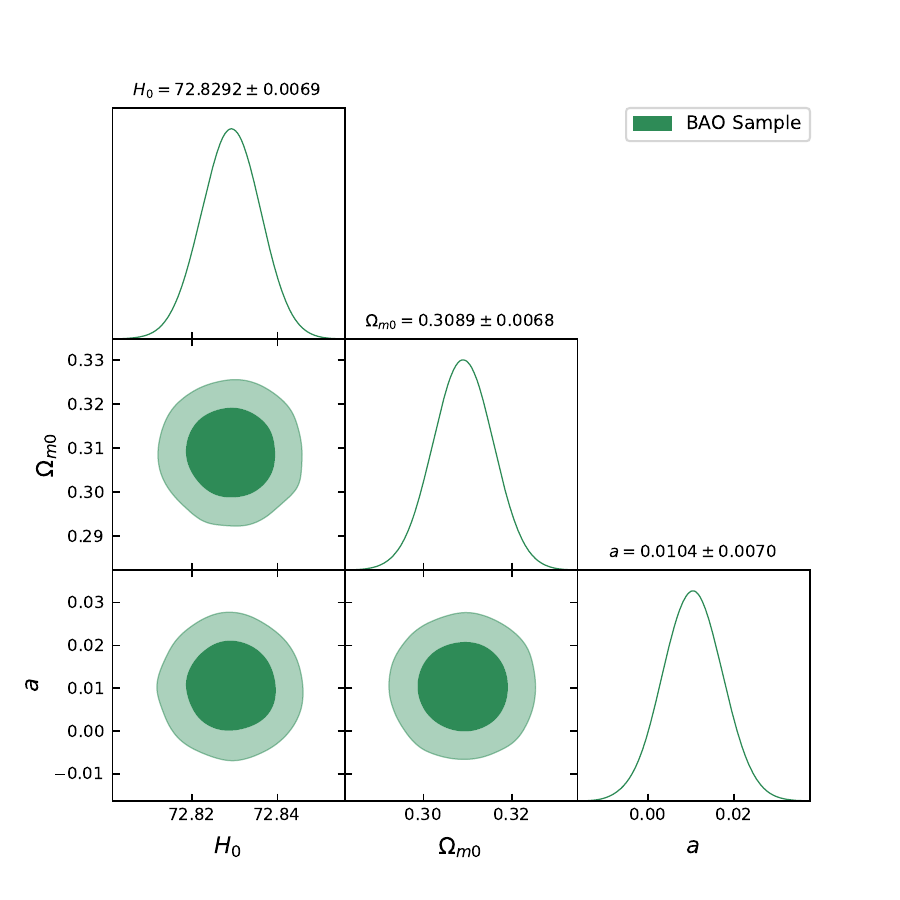}}

\end{figure}

    \begin{table}[H]
\centering
\begin{tabular}{|c ||c c c|} 
 \hline
 & $H_0$ & $\Omega_{m0}$ & $a$ \\ [0.5ex] 
 \hline\hline
 CC & $72.891 \pm 0.0070$ & $0.3106 \pm 0.0067$ & [$0.0028, 0.0168$] \\ 
 Pantheon+SH0ES & $71.7841^{+0.0058}_{-0.0071}$ & $0.2786 \pm 0.0034$ & [$0.0047,0.0185$] \\
 BAO & $72.8292 \pm 0.0069$ & $0.3089 \pm 0.0068$ & [$0.0034,0.0174$] \\
 BBN & ------ & ------ & [$-0.01159,0.01159$] \\[1ex] 
 \hline
\end{tabular}
\caption{Summary of the values of the free parameters obtained from the datasets along with BBN.}
\label{table1}
\end{table}
\end{widetext}

From \autoref{fig6}, we obtain the best-fit ranges of the cosmological parameters $H_0$ and $\Omega_{m0}$ which are in good agreement with the recent observations. Moreover, for model parameter $a$ of the Hybrid model, we find ranges which are overlapping with the range we obtained from the BBN constraints. We have summarized all the obtained values from various constraints in Table-\ref{table1} from which one can find out the common region of $a$ to be [0.00470,0.01159]. This indicates that the model is an excellent alternative to GR because of its efficiency in describing the early time (BBN era), the intermediate time (through the deceleration parameter), and the late time (agreement with different datasets).
\subsection{Statistical comparison of the model with the $\Lambda$CDM model}
To verify the result obtained from the MCMC sample, we perform a statistical comparison of the Hybrid model with the standard $\Lambda$CDM model. The statistical tools Akaike Information Criterion (AIC) and Bayesian Information Criterion (BIC) \cite{Liddle} are utilized for the evaluation. By using the minimum chi-square value obtained from the MCMC, one can obtain the AIC as follows:
\begin{equation}
    AIC=\chi^2_{min}+2d,
\end{equation} 
where d is the number of independent model parameters. Further, the BIC can be defined as
\begin{equation}
    BIC=\chi^2_{min}+d \, lnN,
\end{equation}
where N is the count of data points utilized for the sampling. To compare with $\Lambda$CDM, the difference $\Delta AIC=\abs{{AIC}_{\Lambda CDM}-{AIC}_{MODEL}}$ and $\Delta BIC=\abs{{BIC}_{\Lambda CDM}-{BIC}_{MODEL}}$ are considered. A model is believed to be strongly favored by evidence if $\Delta AIC < 2$, moderately favored if it falls in the range $4 < \Delta AIC \leq 7$, and no significant evidence if $\Delta AIC > 10$. For BIC the ranges can be categorized as $\Delta BIC < 2$ corresponds to strong evidence in favor of the model, $2 \leq \Delta BIC < 6$ indicates the moderate level, and $\Delta BIC > 6$ shows no evidence. We summarize all the obtained quantities from this method in table \ref{table3}.

\begin{widetext}

    \begin{table}[H]
\centering
\begin{tabular}{| c ||c | c | c | c | c|} 
 \hline
  & $\chi^2_{min}$ & AIC & BIC & $ \Delta $AIC & $ \Delta $BIC \\ [0.5ex] 
 \hline
  & Model \hspace{5mm} $ \Lambda $CDM & Model \hspace{5mm} $ \Lambda $CDM & Model \hspace{5mm} $ \Lambda $CDM & & \\ [0.5ex]
 \hline\hline
  CC & 33.5552 \hspace{5mm} 32.1322 & 39.5552\hspace{5mm} 38.1322 & 43.8572\hspace{5mm} 42.4341 & 1.423& 1.423 \\ 
  Pantheon+SH0ES & 1717.228\hspace{3mm} 1609.9172 & 1723.228 \hspace{3mm} 1615.9172 & 1739.542 \hspace{3mm} 1632.2312 & 107.3108 & 107.3108 \\
  BAO & 4.6328 \hspace{7mm} 5.7066 & 10.6328 \hspace{7mm} 11.7066& 10.008 \hspace{5mm} 11.0818 & 1.0738 & 1.0738\\
 \hline
 
 \hline
\end{tabular}

\caption{A collection of values of $ \chi^2_{min} $, AIC and BIC for all the three data sets along with corresponding values for the model and $ \Lambda $CDM.}
\label{table3}
\end{table}
\end{widetext}
 We depict from the AIC and BIC method that the model is strongly favored to compare with the standard $\Lambda$CDM model for the CC and BAO data while it deviates for the Pantheon+SH$0$ES data. 
\section{Conclusion}\label{sec8}
In this work, we have proposed a new functional form for a model in the context of $ f(Q) $ gravity and tested it against both the early and late-time probes available to us. For early-time, we have used BBN constraints on the freeze-out temperature while the late-time study includes the evolution of the Universe as described by the deceleration parameter $ q(z) $. Furthermore, we have used MCMC analysis to obtain the best-fit values of the Hubble parameter $ H_0 $, the density parameter for matter $ \Omega_{m0} $, and finally the model parameter $ a $. The results summarized in table \ref{table1} show that the model parameter range that satisfies all the datasets and BBN constraints is $ a\in [0.00470,0.01159] $. Correspondingly, since $ 3b=a+\Omega_{Q0} $, $ b\in [0.2349,0.2372] $. 

We observe that $ aQ $ is a small correction to the GR equivalent case. Further, the second term $ bQ_0^2/Q $ contributes at lower redshifts when the corresponding value of the Hubble parameter is small since $ Q(z)\propto H^2(z) $. This is synonymous with the effect of dark energy in the $ \Lambda $CDM model in late-time. At early times, however, the second term is significantly suppressed and we are left with a theory that is very close to GR.  

To summarize, the new hybrid model satisfies constraints from BBN in the very early Universe, behaving like GR in that era, after which, through the deceleration parameter, the Universe observes a phase transition from decelerating to accelerating. Throughout the evolution, the Hubble parameter is in excellent agreement with the theoretical model $ \Lambda $CDM and the error bar plots for both Hubble and Pantheon+SH0ES datasets (\autoref{errorbar}). Moreover, the MCMC analysis yields values for $ H_0 $ and $ \Omega_{m0} $ which fits into the currently accepted ranges of the same.

\textbf{Data availability} There are no new data associated with this article.

\acknowledgments  SSM acknowledges the Council of Scientific and Industrial Research (CSIR), Govt. of India for awarding Junior Research fellowship (E-Certificate No.: JUN21C05815).  PKS acknowledges Science and Engineering Research Board, Department of Science and Technology, Government of India for financial support to carry out Research project No.: CRG/2022/001847. We are very much grateful to the honorable referee and to the editor for the
illuminating suggestions that have significantly improved our work in terms
of research quality, and presentation.

\end{document}